\documentclass[11pt]{article}
\usepackage{jcappub}
\usepackage{array}

\usepackage{graphicx}
\usepackage{epsfig}

\newcommand{\fnl}{f_\text{NL}}

\newcommand{\be}{\begin{equation}}
\newcommand{\ee}{\end{equation}}

\newcommand{\lya}{Ly-$\alpha$ forest}
\newcommand{\lyaf}{Ly-$\alpha$ forest}

\title{Bias, redshift space distortions and primordial nongaussianity of nonlinear transformations: application to \lyaf}

\author[a,b,c]{Uro\v s Seljak}

\affiliation[a]{Department of Physics, Department of Astronomy,
and Lawrence Berkeley National Laboratory, University of California, Berkeley}
\affiliation[b]{Institute of Theoretical Physics, University of Zurich}
\affiliation[c]{Institute for the Early Universe, Ewha University, Seoul}

\emailAdd{useljak@berkeley.edu}

\abstract{
On large scales a nonlinear transformation of matter density field 
can be viewed as a biased tracer of the density field 
itself. 
A nonlinear transformation also modifies the redshift space distortions in the same limit, giving rise to a velocity bias.  
In models with primordial nongaussianity a nonlinear transformation generates a scale dependent bias on large scales. 
We derive analytic expressions 
for the large scale bias,
the velocity bias and the redshift space distortion (RSD) parameter $\beta$, as well as the scale dependent bias
from primordial nongaussianity for 
a general nonlinear transformation. 
These biases can be expressed entirely in terms of the one point distribution function (PDF) of the final field and the 
parameters of the transformation. 
The analysis shows that one can view the large scale bias different from unity and primordial nongaussianity bias 
as a consequence of converting higher order correlations in density
into 2-point correlations of its nonlinear transform. Our analysis allows one 
to devise nonlinear transformations with nearly arbitrary bias properties, 
which can be used to increase the signal in the large scale clustering limit. 
We apply the results to the ionizing equilibrium model of 
Lyman-$\alpha$ forest, in which Lyman-$\alpha$ flux $F$ is related to the density perturbation $\delta$ via a 
nonlinear transformation. 
Velocity bias can be expressed as an average over the Lyman-$\alpha$ flux PDF. 
At $z=2.4$ we predict the velocity 
bias of -0.1, compared to the observed value of $-0.13 \pm 0.03$. 
Bias and primordial nongaussianity bias depend on the parameters of the transformation. 
Measurements of bias can thus be used to constrain these parameters, and  
for reasonable values of the ionizing background intensity we can 
match the predictions to observations. Matching to the observed values we predict the ratio of primordial 
nongaussianity bias to bias to have the opposite sign 
and lower magnitude than the corresponding values for the highly biased galaxies, 
but this depends on the model parameters and can also vanish or change the sign. 
}

\begin{document}
\maketitle
\flushbottom

\section{Introduction}

A simple model for \lyaf\ 
relates the neutral hydrogen responsible for absorption to the underlying matter density field 
via a chemical equilibrium equation, 
where recombinations and photo-ionizations balance each other \cite{1998ApJ...495...44C}. 
Optical depth is 
proportional to neutral hydrogen density, which in the ionizing equilibrium can be related to gas density as $\tau=A(1+\delta)^{\alpha}$
\footnote{In some previous papers $\beta$ is used in place of $\alpha$, but here we will reserve $\beta$ for the redshift 
space distortion parameter.}. 
Here $\delta$ is the gas overdensity parameter, 
$\alpha=2-0.7(\gamma-1)$, where $\gamma-1=d\ln \rho /d\ln T$ is the slope of temperature-density relation 
and the recombination coefficient is assumed to scale as $T^{-0.7}$. 
Typical value is $\alpha=1.6$, with $\alpha=2$ being the isothermal case. 
The observed flux $F$
is related to the
optical depth $\tau$ as $F=\exp(-\tau)$. 
The relation between the observed flux and the density perturbation $\delta$ is thus highly noninear. 

Even though the relation between the \lya\ and the underlying matter density is nonlinear, on large scales \lya\ fluctuations
trace the dark matter fluctuations up to a constant factor of proportionality called density bias.
The basic premise of Lyman-$\alpha$ forest clustering analyses is that the bias is a known function of the underlying parameters. 
This is the basis of the statements that \lya\ measures directly the amplitude of matter fluctuations at $2<z<4$ \cite{2005ApJ...635..761M}. 
However, so far all of the predictions came from simulations and 
we do not have a good analytic understanding of how the bias in \lya\ is determined by the parameters of the model. 
The purpose of this paper is to derive 
the bias analytically and to explore its sensitivity to the 
physical parameters. 
We will assume the gas density is related to the dark matter density 
smoothed on the Jeans scale (or, more precisely, filtering length, \cite{1998MNRAS.296...44G}). 
This relation is not exact and there is scatter around it, at the level of 10-50\% \cite{1997MNRAS.292...27H}. 
Initially we assume the relation between optical depth and matter density is deterministic, later we generalize this 
to include a simple form of scatter. 

Our second motivation is to derive an analytic prediction for the redshift space distortions (RSD). 
In redshift space the observed position is a sum of the radial distance (in velocity units) and the radial velocity. 
The velocity gradients give rise to additional perturbations called RSD. 
One can view this as a mapping from the real space to the redshift space with the 
total number of tracers being conserved. 
In the large scale limit 
the RSD take a particularly simple form first derived by Kaiser \cite{1987MNRAS.227....1K}. In this limit the large scale 
velocity of any tracer follows the dark matter and there is no velocity bias. 
However, if the field is transformed {\it after} the RSD mapping then the transformed field acquires a velocity bias different from unity. 
This is the case for \lya\, since RSD act on optical depth $\tau$, 
while the observable is $F=\exp(-\tau)$. In this paper we derive its velocity bias. Combining with the bias predictions one 
can also predict the RSD parameter $\beta$, which is defined as the ratio between the velocity bias and the density bias. 
Recent observations of bias and 
$\beta$ provide an opportunity to compare our predictions to the observations \cite{2011JCAP...09..001S}. 

The third motivation for this paper is to explore the sensitivity of \lya, and nonlinear transforms in general, 
to the primordial nongaussianity. Primordial 
nongaussianity models have an additional contribution added to the primordial density field, which
does not show up in its 2-point correlation function, but only in the higher order correlations. 
However, such a component can show up in the 2-point correlations of the nonlinear transform of the density field. 
We derive its amplitude and scale dependence 
and show that it agrees with the 
corresponding scale dependent bias  of biased halos \cite{2008PhRvD..77l3514D,2008ApJ...677L..77M,2008JCAP...08..031S} 
up to a prefactor which is determined by the transformation parameters. 

Although the \lya\
will be our primary application in this paper, the formalism we develop here 
is more general and can be applied to any tracer of the dark matter. 
The goal of this work is to investigate the large scale density bias (hereafter bias), velocity bias (as defined in 
redshift space distortions) 
and scale dependent primordial nongaussianity bias 
of a general nonlinear transformation 
of the matter density field $\delta$. 
For example, galaxies are formed inside dark matter halos and these are often modeled as a nonlinear transform
of the local density field, $\delta_h=b_1\delta+b_2\delta^2+...$ \cite{1993ApJ...413..447F}, where density is  
smoothed on a scale typically related to the Lagrangian scale of galaxy sized halos (which is of order 1 Megaparsec). 
The linear bias $b_1$ is a monotonically increasing function of mass of the halos in which 
the galaxies live. This is usually explained within the context of a universal halo mass function and the peak background split 
\cite{1996MNRAS.282..347M}. 
One can however also view the bias different from unity as a consequence of a nonlinear transformation of the density field: 
as we will show in this paper a nonlinear transformation can result in an arbitrary value of bias.
This analogy is not perfect: galaxies and halos are discrete objects and in addition to the nonlinear transformation one must also impose 
the exclusion constraint, where no halos can be within the virial radius of each other. 
However, one can also consider further nonlinear transforms of the galaxy field which modify the bias properties. This may be useful 
for observations where individual galaxies are not measured, only their overall intensity imprints, such as in the 21-cm 
intensity mapping \cite{2008PhRvL.100i1303C}. 

\section{Formalism}

We begin with the derivation of the large scale bias. 
Let us call the nonlinear transformation of the density field $\tau(\delta)$ and
decompose the density perturbation into a long wavelength component $\delta_l$
and a short wavelength component $\delta_s$, $\delta=\delta_l+\delta_s$, both with zero average. 
We will also assume $|\delta_l| \ll |\delta_s|$ in an rms sense. This assumption is well justified
in our universe where power per mode $k^3P(k)$ is rapidly increasing with wavevector $k$ ($P(k)$ is the power spectrum). 

We want to know the response of the nonlinear transform $\tau(\delta)$
to a long wavelength mode $\delta_l$. 
We define the 
(density) bias as
\be
b_{\tau}=\left\langle {\partial \tau \over \partial \delta_l} \right\rangle. 
\ee
Here $\langle \rangle$ denotes average over the field. 
We can expand $\tau(\delta)$ in a Taylor series
\be
\tau(\delta)=\sum_{n=0}^{\infty} {\tau^{(n)}(0)\delta^n \over n!},
\ee
where $\tau^{(n)}(0)$ is the $n$-th derivative of function $\tau$ evaluated at $\delta=0$. 
The first term 
in expansion above is a constant and does not depend on $\delta_l$. The second term, $n=1$, is linear 
in $\delta_l$, so its dependence on $\delta_l$ is simply $\tau^{(1)}(0)\delta_l$ and the first 
order bias of $\tau$ is $b_{\tau}^1=\tau^{(1)}(0)$. 

To understand the bias for higher order terms ($n>1$), 
we need to develop understanding of gravitational coupling between the long wavelength and the short wavelength modes. 
This can be achieved using perturbation theory, but here we will pursue a simpler approach in terms of a 
constant overdensity $\delta_l$. A similar derivation has recently been given in \cite{2011JCAP...10..031B}. 

\subsection{Coupling between long and short wavelength modes}

We would like to analyze the response of small scale perturbations to a long wavelength 
density perturbation $\delta_l$, which we will model as constant in space. This leads to a locally slightly overdense
or underdense universe, but one where the global time and coordinates
are still given by the global unperturbed value. 
Thus one must consider a slightly changed Hubble expansion rate, as well as the fact that the 
small scale density perturbation feels the additional gravity from the overdensity (or underdensity) of the 
long wavelength mode. 
In addition, the long wavelength overdensity itself grows in time according to the linear growth rate. 
Most of the time during matter domination the universe is EdS with $\Omega_m=1$, so we 
focus on that solution.  We write the expressions in terms of an overdensity perturbation $\delta_l>0$, although the
final result is the same for an underdensity $\delta_l<0$. 

We begin with the standard approach in deriving the spherical collapse model solution. 
We work in the locally comoving coordinates, but using global time $t$. 
In a homogeneous universe the cycloid solution to the expansion is 
\be
{a \over a_m}={1 \over 2} \left(1-\cos\eta \right),\\\\\\\\{t \over t_m}={1 \over \pi}\left(\eta-\sin \eta \right)
.
\ee
Taylor expanding in small $\eta$ and eliminating $\eta$ from the expressions order by order 
consistently to get the first two non-vanishing terms gives 
\be
a=\left({t \over t_0 }\right)^{2/3}\left(1-{1 \over 3}t^{2/3}\delta_{l0}\right).
\ee
Here time $t_0$ is the age of the EdS universe today, and the factors have been arranged such 
that the long wavelength mode is $\delta_l=\delta (a^{-3})=t^{2/3}\delta_{l0}$, 
i.e. the long wavelength mode is growing according to its linear growth rate, which in an 
EdS universe is just the expansion rate $a=(t/t_0)^{2/3}$. We have applied mass conservation, 
so the long wavelength density perturbation is given simply by the change in volume $\delta (a^{-3})$. 
The corresponding Hubble rate to the same order is 
\be
{\dot{a} \over a}={2 \over 3t}\left(1-{1 \over 3}t^{2/3}\delta_{l0} \right).
\ee

A short wavelength perturbation $\delta_s$ in this universe obeys the equation 
\be
\ddot{\delta_s}+2{\dot{a} \over a}\dot{\delta_s}=4\pi G \bar{\rho}\delta_s\left(1+t^{2/3}\delta_{l0}\right).
\label{ds}
\ee
This is the usual second order equation derived from continuity and Euler's equation for dark matter. 
The last term includes the fact that gravity responds to the total matter and so 
in a slightly overdense region (caused by $\delta_l$) the gravitational force will 
be slightly stronger. Note that this term vanishes initially, at $t=0$, so it does not enter
in setting up the initial conditions and there is no coupling between the modes initially, as expected. 
Here $\bar{\rho}$ is the global density and we have $H^2=8\pi G \bar{\rho}/3=(2/3t)^2$. 
We know that for $\delta_0=0$ the solution is $\delta_s=\delta_{s0}t^{2/3}$, so we
can write the ansatz solution as $\delta_s=\delta_{s0}t^{2/3}(1+\beta_2 t^{2/3}\delta_{l0})$. 
Inserting this ansatz into equation \ref{ds} gives $\beta_2=13/21$. 

This solution was with respect to the local expansion, i.e. wrt to the local comoving coordinates. With respect to 
the global coordinates we have 
\be
\delta_s^g=(1+\delta_l)\delta_s=(1+\delta_l)(1+\beta_2 \delta_l)a\delta_{s0}\sim 
\left(1+{34 \over 21} \delta_l\right) a\delta_{s0} \equiv \left(1+\nu_2 \delta_l\right) a\delta_{s0}.
\label{nu2}
\ee
We thus derived the result that the short scale density perturbation is enhanced
in the presence of a long wavelength perturbation $\delta_l$ by a multiplicative factor 
proportional to $\delta_l$. 
The coefficient of proportionality
$\nu_2=34/21$ is the well known angular average of the second order perturbation 
theory kernel $F_2({\bf k_1},{\bf k_2})$ \cite{2002PhR...367....1B}. The small scale perturbations are also 
rescaled in size by $\delta_l/3$, but this effect is not relevant for 
the purpose of this paper. 

\subsection{Large scale bias}

With this result in hand we can write to the lowest order in $\delta_l$, 
\be
\delta^n=(\delta_s+\delta_l)^n=[\delta_s(1+\nu_2\delta_l)+\delta_l]^n \sim \delta_s^n(1+n\nu_2 \delta_l)+n\delta_s^{n-1}\delta_l. 
\ee
Thus
\be
b_{\tau}={\partial \tau \over \partial \delta_l}=
\sum_{n=1}^{\infty} {\tau^{(n)}(0) \over n!}{\partial \delta^n \over \partial \delta_l}= \\
\nu_2 \sum_{n=2}^{\infty} n{\tau^{(n)}(0)\delta^n \over n!}+
\sum_{n=1}^{\infty} n{\tau^{(n)}(0)\delta^{n-1} \over n!} = \\
\nu_2\left\langle \delta {d\tau\over d\delta} \right\rangle+\left\langle {d\tau \over d\delta} \right\rangle.
\ee
We used $\delta \sim \delta_s$ at the lowest order. 

The simplest non-trivial example is that of a quadratic dependence, such as the optical depth of 
Lyman-$\alpha$ forest in the case of isothermal density-temperature relation, $\tau=A(1+\delta)^2$. 
We find 
$b_{\tau}=2A(1+\nu_2\sigma_J^2)$, 
where $\sigma_J$ is the rms density field smoothed on a 
Jeans scale. 
Note that we can get the variance from the transformed field itself, i.e. $\sigma_J^2=A^{-1}\langle \tau \rangle-1$.
A more meaningful way to express this is to look at the bias of 
optical depth overdensity $\delta \tau \equiv \tau/\langle \tau \rangle-1$, 
\be
b_{\delta \tau}={2(1+\nu_2\sigma_J^2) \over 1+\sigma_J^2}. 
\label{btau}
\ee
In the limit of small $\sigma_J^2$ this agrees with the linearized analysis, $b_{\delta \tau}=2$, 
 where only the first term in Taylor expansion is kept. On the other hand, in the limit of large $\sigma_J^2$ we get 
$b_{\delta \tau}=2\nu_2=3.24$.
In the case of Lyman-$\alpha$ forest applications, for $2<z<4$, $\sigma_J^2$ is expected to be of order unity, 
which means that the bias contribution from the quadratic term 
is important relative to the linear term. 
The derivation assumed that the short wavelength modes grow according to linear theory, so 
this approximation presumably breaks down if the variance of the density field $\sigma_J \gg 1$. 
Limited tests in simulations however suggest that this approximation works well even if $\sigma_J \gg 1$, 
e.g. figure 1 of \cite{2011PhRvD..84h3509H}. 

\subsection{Primordial nongaussianity}

The case of primordial nongaussianity of local type is even simpler. The local model for 
initial potential is
\be
\Phi_\text{nG}(\vec x)=\varphi(\vec x)+\fnl\left(\varphi^2(\vec
x)-\left\langle\varphi^2\right\rangle\right)
,\label{eq:ngpot}
\ee
where $\varphi$ is primordial Gaussian potential. 
Performing the same long-short wavelength mode split as above one finds, 
\be
\delta_{s,\text{nG}}=\delta_s(1+2\fnl \alpha_{\fnl}^{-1}\delta_l),
\label{fnl}
\ee
where the relation between initial potential and final linear density is given by
\be
\alpha_{\fnl}(k,z)=\frac{2 k^2 c^2 D(z) T(k)}{3H_0^2
\Omega_\text{m}},\label{eq:alphaparam}
\ee
where $H_0$ is the Hubble parameter, $T(k)$ the transfer function an $D(z)$ the growth rate. 
Note that $\alpha_{\fnl}$ scales as $k^2$ on large scales where the transfer function is unity, 
hence the effect becomes large on very large scales.

We see that in equation \ref{fnl} $2\fnl \alpha_{\fnl}^{-1}$ replaces $\nu_2$ in equation \ref{nu2}.
The large scale bias due to the nongaussian term is thus
\be
b_{\tau,NG}=2\fnl \alpha_{\fnl}^{-1}\left\langle \delta {d\tau\over d\delta} \right\rangle .
\ee
Other primordial nongaussianity model can be treated the same way, except that 
$\alpha_{\fnl}$ changes, e.g. it scales as $k^{-1}$ for orthonormal models and 
is constant for equilateral models \cite{2010JCAP...01..028S}. 

\subsection{Redshift space distortions}

A third application are the redshift space distortions(RSD). 
The situation we wish to consider is one where observations are in redshift space, meaning the 
position of the object (or intensity for continous case) has to include the peculiar velocity of the object. 
There are two cases that can be considered. One is when 
RSD transformation occurs after the nonlinear transformation. 
In this case RSD simply remaps the variable $\tau$ from the real space to the redshift space, such that the total is 
conserved, i.e. $\tau({\bf r})d^3{\bf r}=\tau({\bf s})d^3{\bf s}$, where $\bf{r}$ denotes real space coordinate
and $\bf{s}$ the redshift space counterpart. The Jacobian of the transformation is $|d^3{\bf s}/d^3{\bf r}|=1-dv_z/dz$, 
where we denote with $z$ the radial direction and $v_z$ is the velocity in the radial direction. 
If we consider the response to a Fourier mode $\delta_l$  
then the usual linear order result gives
$dv_z/dz=f\mu^2\delta_l$, where $\mu$ is the
angle between the Fourier mode direction and line of sight and $f$ is the logarithmic growth rate \cite{1987MNRAS.227....1K}. At 
linear order we thus have
\be
\tau({\bf s}) = \tau({\bf r})(1+f\mu^2\delta_l).
\ee
we thus find that the lowest order contribution from RSD is $\langle \tau \rangle f\mu^2\delta_l$.
Note that if the field is divided by the mean then there is no velocity 
bias in RSD, i.e. velocities are a faithful tracer of the matter field, which is the usual result 
for galaxies. 

A more complicated case is where the nonlinear transformation occurs after RSD mapping. 
An example is the flux in Lyman-$\alpha$ forest. The optical depth $\tau$ is proportional to 
the neutral hydrogen along the line of sight, which is remapped into the redshift space due 
to peculiar velocities, giving optical depth $\tau(\bf{s})$. The observable however is the 
flux $F$ relative to unabsorbed value (continuum), i.e. the fraction of the flux 
absorbed by the neutral hydrogen is $F[\tau({\bf s})]=\exp[-\tau({\bf s})]$. 
To preserve generality we will however assume $F(\tau)({\bf s})$ is a general function of $\tau$. 
We can expand in $\tau$, and at the lowest order in $\delta_l$ we have
\be
F[\tau({\bf s})]=F[\tau({\bf r})(1+f\mu^2\delta_l)]=\sum_{n=0}^{\infty} {F^{(n)}(0)\tau^n(1+nf\mu^2\delta_l) \over n!}
=F[\tau({\bf r})]+f\mu^2 \delta_l \left\langle \tau {dF \over d \tau} \right\rangle. 
\ee

Combining all previous results we thus arrive at the final expression for the bias, 
\be
b_F(\mu,\fnl)={\partial F[\tau(\delta({\bf s}))] \over \partial \delta_l}=\left(\nu_2+2\fnl \alpha_{\fnl}^{-1} \right) \left\langle \delta {d F \over d \delta} \right\rangle 
+ \left\langle {d F \over d \delta} \right\rangle +f\mu^2 \left\langle \tau{dF \over d\tau}\right\rangle.
\ee
This is the central equation of this paper. It has a remarkable property that if 
the nonlinear transform is monotonic we can invert the relations and express everything in terms of one point distribution function (PDF) of 
the final observable $F$ and the parameters of the nonlinear transformation.  
Often the fluctuating field is normalized to unity, i.e. the field is defined to be 
$\delta F \equiv F({\bf s})/\langle F \rangle-1$. In this case the bias above is divided by $\langle F \rangle$.

To simplify the notation we can define the (density) bias of $F$ as \be
b_F=\nu_2\left\langle \delta {d F \over d \delta} \right\rangle
+ \left\langle {d F \over d \delta} \right\rangle,
\ee
the primordial nongaussianity bias as 
\be
b_{\fnl}=2\left\langle \delta {d F \over d \delta} \right\rangle
\label{bfnl}
\ee
and velocity bias of $F$ as
\be
b_v=\left\langle \tau{dF \over d\tau}\right\rangle,
\ee
such that 
\be
b_F(\mu,\fnl)=b_F+b_{\fnl}\fnl \alpha_{\fnl}^{-1}+fb_v\mu^2.
\ee
It is customary to introduce the RSD parameter $\beta$ as $b_F(\fnl=0)=b_F(\mu=0)(1+\beta \mu^2)$, in which case
\be
\beta_F \equiv {fb_v \over b_F}={f\langle \tau{dF \over d\tau}\rangle \over \nu_2 \langle \delta {d F \over d \delta} \rangle
+ \langle {d F \over d \delta} \rangle}.
\ee

\subsection{Galaxies and nonlinear transformations}

We have shown above that a nonlinear transformation of the density field changes the bias. 
This happens even if there is no bias at the linear order, i.e. in $\delta_h=b_1\delta+b_2 \delta^2...$ the 
unrenormalized value is $b_1=1$. 
It is well known that dark matter halos in which galaxies form have bias different from unity, 
ranging from 0.7 at the low mass end to an arbitrarily large value at the high mass end
\cite{1996MNRAS.282..347M}. In the picture pursued in this paper galaxies and 
halos with $b \ne 1$ can be viewed as a result of a nonlinear transformation. 
In this picture the bias different from unity happens because 
higher order correlations in 
$\delta$ show up as 2-point correlations in the nonlinear transform of $\delta$, hence the 
large scale bias is renormalized \cite{2006PhRvD..74j3512M,2008PhRvD..78l3519M}. 
So one can ask how far can this analogy be pursued. 
For example, we have seen 
that for $\tau=\delta^n$ one gets $b_{\delta \tau}=n\nu_2$ if $n$ is even, so one can obtain an 
arbitrarily large bias from such transforms. One can lower this to an arbitrary low number, 
for example by adding a constant component, e.g. for $\tau=A+\delta^n$ one has 
\be
b_{\delta \tau}={n\nu_2 \over A+\langle \delta^n \rangle},
\ee
for even $n$. 
As we have seen in previous section in the application to 
\lya\, one can also easily obtain $|b|<1$. 

Primordial nongaussianity also leads to a scale dependent bias: for $\tau=\delta^n$ we find 
$b_{\fnl}=2n$, so we have 
\be
{b_{\fnl} \over b_{\delta \tau}}={2 \over \nu_2} \sim 1.2. 
\label{fnldn}
\ee
If there is a linear component added to this with a positive sign, e.g. $\tau=\delta^n+A\delta$ and $A>0$, 
then this number will be reduced, because a linear component adds to the 
bias $b_{\delta \tau}$ but not to the primordial nongaussianity bias. Of course,
if only the linear component is present 
then $b_{\fnl} =0$. 
The ratio $b_{\fnl}/b$ can also be increased 
if we consider $A<0$, which reduces $b_{\delta \tau}$ but not 
$b_{\fnl}$. 
As discussed above this can be compared to 
${b_{\fnl} \over b}=2\delta_c(b-1)/b \sim 3.37(b-1)/b$, where $\delta_c=1.68$
for biased halos \cite{2008PhRvD..77l3514D}. 
We thus find that a nonlinear transform can accommodate the complete range of allowed values of bias and 
primordial nongaussianity bias.

A few more remarks are in order. First, if these nonlinear transforms act on the density field and the 
redshift space distortions act on the nonlinear transform then there is no velocity bias, consistent 
with the standard picture of halos and galaxies. 
Second, halos are discrete objects and in addition to the nonlinear transformation one must also impose
the exclusion constraint, where no halos can be within the virial radius of each other.
This constraint cannot be expressed in terms of just the local density transformation (the situation is even more
complicated for real galaxies, which can be either central galaxies or satellites in a halo).
Third, even if the nonlinear transform can give an arbitrary value of the bias, the corresponding shot noise 
can differ significantly from that of the discrete halo tracers. So in detail there are considerable differences
between the real galaxies or halos and the nonlinear transform of the density field, even if both display large 
scale bias and primordial nongaussianity bias. 

One can also consider further nonlinear transforms 
of the galaxy density field, which will result in a bias different from the original field. For example, this way 
one can construct a tracer with a different large scale bias than the original galaxy field, which may be useful 
if one wants to apply the multiple tracer sampling variance methods as in \cite{2009JCAP...10..007M,2009PhRvL.102b1302S}. 
One such application where this 
could be useful is 21-cm intensity mapping \cite{2008PhRvL.100i1303C}. 

\section{Application to Lyman-$\alpha$ forest}

The above derivations are entirely general and can in principle be applied to any 
nonlinearly transformed density field. 
Let us apply the above results to the model for Lyman-$\alpha$ forest, ignoring the primordial nongaussianity for the moment. 
The optical depth is given by $\tau=A(1+\delta)^{\alpha}$ and the flux is $F=\exp(-\tau)$. 
This gives 
\be
b_F=-A\alpha\nu_2\left\langle (1+\delta)^{\alpha-1}\delta F \right\rangle-A\alpha \left\langle (1+\delta)^{\alpha-1} F \right\rangle.
\ee
This can be rewritten as 
\be
b_F=-A\alpha(\nu_2-1)\left\langle (1+\delta)^{\alpha-1}\delta F  \right\rangle-A\alpha \left\langle (1+\delta)^{\alpha} F \right\rangle.
\ee

The velocity bias is 
\be
b_v=-A\left\langle (1+\delta)^{\alpha} F \right\rangle,
\ee

The redshift space distortion parameter $\beta_F$ is
\be
f\beta_F^{-1}=\alpha+ (\nu_2-1){\left\langle \delta {dF \over d\delta} \right\rangle \over \left\langle \tau {dF \over d\tau} \right\rangle}. 
\ee
Note that for $z>2$ we can approximate $f \sim 1$ to a high accuracy and we will assume $f=1$ below. 
In the expression for $\beta_F$ above, the term $\left\langle (1+\delta)^{\alpha} F  \right\rangle$ is 
always positive, while $\left\langle \delta (1+\delta)^{\alpha-1} \delta F \right\rangle$ can be of either sign 
because $\delta$ is of either sign. 
As shown below it is typically negative, and as a result the RSD parameter is typically larger than $\alpha^{-1}$. 

Since the relation between matter density and flux is monotonic we can invert the above relation, i.e. 
we can write $\delta(F)=(-\ln F/A)^{\alpha^{-1}}-1$. This way  we obtain 
an expression that only depends on the parameters of the nonlinear transformation, $A$ 
and $\alpha$, and on the one point distribution function (PDF) of the final field $F$ itself. 
In terms of flux and its PDF bias can be expressed as
\be
b_F=\alpha \langle F \ln F \rangle +\alpha(\nu_2-1)\left\langle F \ln F [1- (-\ln F/A)^{-\alpha^{-1}}]\right\rangle,
\label{bFF}
\ee
and
\be
b_v=\langle F \ln F \rangle.
\label{bFv}
\ee
We can thus use the physical model parameters and the observed flux PDF to determine the bias, without 
any need for simulations. 
Note that velocity bias $b_v$ is completely independent of the parameters of the nonlinear transformation and is
negative definite across the entire range of $F$, hence it is always negative. 
In contrast, $b_F$ explicitly depends on $A$ and $\alpha$, so these 
two parameters must be specified in addition to flux PDF. Moreover, the last term contains $\langle F (1+\delta)^{\alpha-1} \rangle$, which 
becomes comparably large in the voids where $\delta \sim -1$, and this terms contributes with opposite sign to the first two terms 
in equation \ref{bFF}. The predictions for $b_F$ are thus very sensitive to the void regions. 

\subsection{Log-normal models predictions}

To proceed we must evaluate these terms. We could simply use the observed PDF of $F$ and 
explore the predictions as a function of $A$ and $\alpha$ and we will do so in the next section. However, we also want to 
develop a better analytic understanding of where the dominant 
contributions come from, so we will use the log-normal model, which we show to 
give a reasonable approximation to the PDF of the flux field. In this 
model the nonlinear density field is given by $1+\delta=\exp(\delta_G-\sigma^2/2)$, 
where $\delta_G$ is a gaussian random field, with a gaussian probability distribution 
$p(\delta_G|\sigma)= (2\pi \sigma^2)^{-1/2}\exp[-{\delta_G^2 \over 2\sigma^2}]$. 
Note that the variance $\sigma$ is related to the Jeans smoothed variance of density field, $\sigma_J^2=\exp(\sigma^2)-1$. 
For a given value of $\alpha$
the constants $A$ and $\sigma$ can be determined by normalizing to the observed mean flux and rms variance of flux. 
The mean flux is  $\bar{F}=0.47 \pm 0.02$ at $z=4$, $\bar{F}=0.68 \pm 0.02$ at $z=3$ and 
$\bar{F}=0.82 \pm 0.01$ at $z=2.4$, while $\sigma_F^2=0.013 \pm 0.003$ at $z=4$, 
$0.0117 \pm 0.006$ at $z=3$ and $0.0079 \pm 0.007$ at $z=2.4$ \cite{2002ApJ...580...42M}. 

With this the model is fully specified and the bias and $\beta$ can be calculated by doing 
two simple gaussian integrals. These can be written as 
\be
\langle (1+\delta)^{\alpha-1}\delta F \rangle 
=(2\pi \sigma^2)^{-1/2}\int e^{(\alpha-1)(\delta_G-\sigma^2/2)}\left(e^{\delta_G-\sigma^2/2}-1\right)e^{-Ae^{\alpha(\delta_G-\sigma^2/2)}}e^{-{\delta_G^2 \over 2\sigma^2}}
d\delta_G
\ee
and 
\be
b_v=\langle (1+\delta)^{\alpha}F\rangle =(2\pi \sigma^2)^{-1/2}\int e^{\alpha(\delta_G-\sigma^2/2)}
e^{-Ae^{\alpha(\delta_G-\sigma^2/2)}}e^{-{\delta_G^2 \over 2\sigma^2}}d\delta_G. 
\ee

These integrals can be written in a simpler form by transformation $dF/d\delta_G=-A\alpha F(1+\delta)^{\alpha}$, 
\be
b_F=-\int_0^1 dF p(\delta_G|\sigma)\left[{\nu_2\delta +1 \over 1+\delta} \right]
\ee
and
\be
b_v=-\alpha^{-1}\int_0^1 dF p(\delta_G|\sigma),
\ee
and we note that $\delta_G$ is an implicit function of $F$ whose explicit form can be obtained
by inverting the relation $F=\exp[-Ae^{\alpha(\delta_G-\sigma^2/2)}]$. 

Finally, for a given $\alpha$ 
the parameters $A$ and $\sigma$ can be determined from the mean flux and rms variance, which are given by 
\be
\langle F \rangle = \int_0^1 dF p(\delta_G|\sigma) { 1 \over A\alpha \tau}
\label{meanf}
\ee and
\be
\langle F^2 \rangle = \int_0^1 dF p(\delta_G|\sigma) { F \over A\alpha \tau},
\label{meanf2}
\ee
with $\sigma_F^2=\langle F^2 \rangle-\langle F \rangle^2$, or
\be
\sigma_F^2=\int_0^1 dF p(\delta_G|\sigma)\left[{(F-\langle F \rangle)^2 \over A\alpha \tau F}\right].
\label{sigmaf2}
\ee

We can write these equations above as 
\be
X=\int dF W_X(F),
\ee
where $W_X$ is the integration kernel and $X=b_F,\, b_v,\, F,\, \sigma^2_F$ and can be either taken from the 
lognormal model or directly expressed from the PDF, as in equations \ref{bFF}-\ref{bFv}. 
These have a simple interpretation: the measured value of $X$ is simply the average of the 
integration kernel over the interval $0<F<1$.

\begin{figure}[t!]
\begin{center}
 \vspace{-10mm} \hspace{18mm}
 \resizebox{1.00\hsize}{!}{\includegraphics{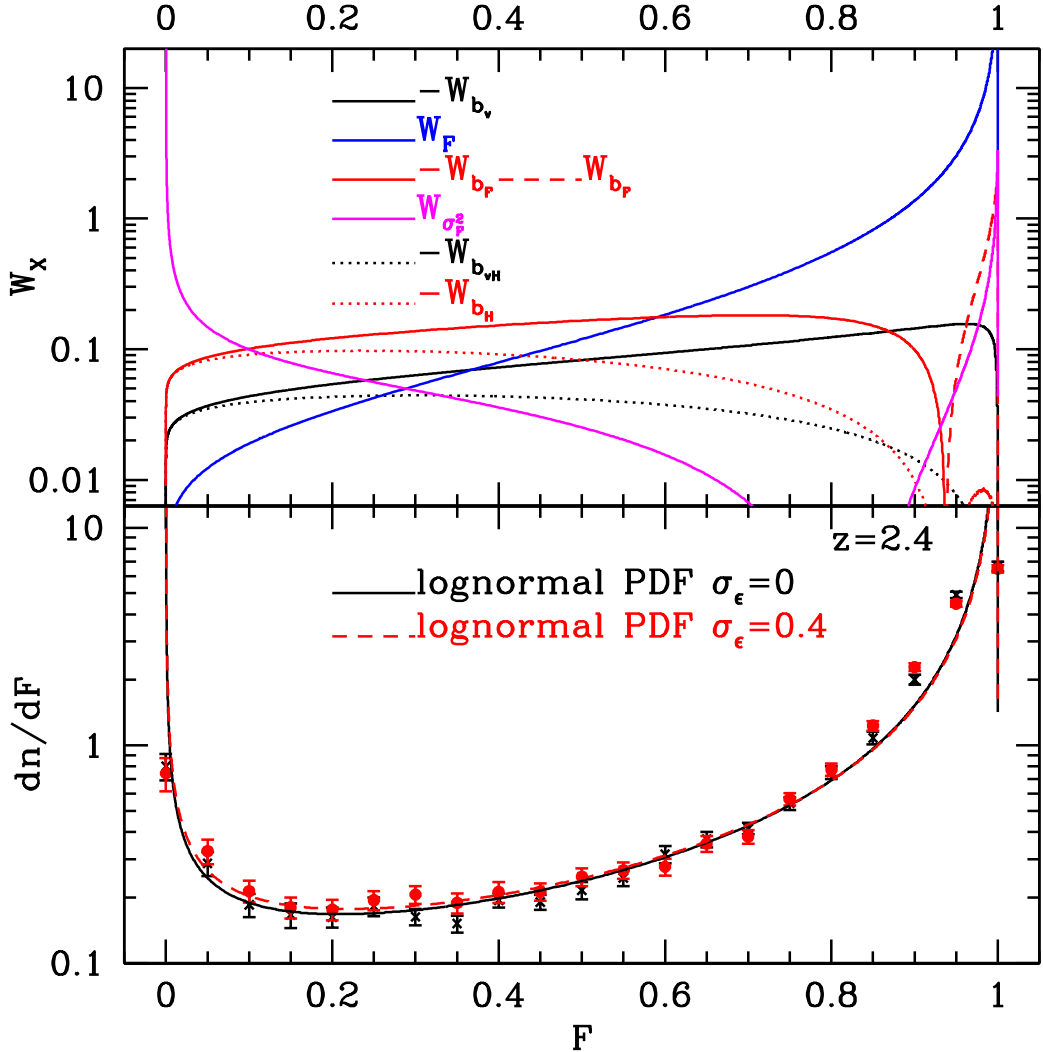}} \\
 \vspace{-40mm}
\end{center}
\caption{
Top panel shows window functions $W_X$ for bias ($W_{b_F}$) and velocity bias ($b_{vF}$) of $F$, 
as well as for mean flux ($W_{F}$) and flux rms ($W_{\sigma_F^2}$) for a representative log-normal model ($\sigma=1.5$, $\alpha=1.6$) at $z=2.4$. 
The window has the same sign for velocity bias $b_{vF}$ over the entire interval, 
while the window for bias $W_{b_F}$ crosses zero and becomes large for $F>0.93$, suggesting a large 
contribution from void regions. We also show window functions of $H=F-F^2/2$,
both bias ($W_{b_H}$) and velocity bias ($b_{vH}$), for which the void contribution is suppressed. Bottom figure shows the PDF from this model, 
compared to observations in \cite{2001ApJ...549L..11M,2007MNRAS.382.1657K}, indicating that log-normal model provides a decent fit to the data. 
We also show the deconvolved PDF of $dn/d\bar{F}$ derived from the log-normal PDF of $dn/dF$, 
assuming a scatter of $\sigma_{\epsilon}=0.4$, as described in the text. 
We see that this amount of scatter has a relatively modest effect on the PDF. 
}
\label{fig1}
\end{figure}

For $z=2.4$, we find that $\sigma=1.5$, $A=0.3$ 
and $\alpha=1.6$, give a reasonably good fit to the flux PDF, 
as shown in the bottom of figure \ref{fig1}. The corresponding mean flux and rms are
$\langle F \rangle=0.82$ and $\sigma_F^2=0.07$, in good agreement with observed values \cite{2002ApJ...580...42M}. 
The predicted values of bias are $b_v=-0.1$ and $b_F=-0.11$. 
At higher redshifts the agreement between observed flux PDF and log-normal model becomes worse, so log-normal 
model is less useful there and we do not show the results here. 

It is useful to understand what part of the flux the
dominant contributions come from within the log-normal model. 
The integration kernels are shown in top of figure \ref{fig1}. 
One can see that the mean flux is heavily dominated by the high $F$ region, as expected, caused
by the presence of $\tau$ in the denominator of $W_F$ kernel, which leads to divergence
for $\tau=0$, $F=1$. This divergence is cured by the gaussian PDF $p(\delta_G|\sigma)$, but this 
happens only for values extremely close to $F=1$, so the preference towards high values of $F$ remains. 
This divergence in itself is not very meaningful: we could also have 
defined the relevant quantity as $\langle 1-F \rangle$, which would have cured the divergence at $F=1$. 

The flux rms $\sigma_F^2$ receives contributions away from the mean flux since 
$W_{\sigma_F^2}$ kernel contains $(F-\langle F \rangle)^2$ term. It is again 
divergent at $F=1$ because of $\tau=0$ in the denominator, but is also divergent at $F=0$ 
because of $F$ in the denominator of equation \ref{sigmaf2}
(this is again cured by the gaussian PDF $p(\delta_G|\sigma)$, but this 
only happens very close to $F=0$). 
The peak 
at low $F$ is significantly higher than the one at high $F$. 
The rms fluctuations are thus
heavily dominated by the high absorption regions very close to $F=0$. 

In contrast to $\langle F \rangle$ and $\sigma_F^2$, 
the bias $b_F$ and velocity bias $b_v$ are 
more broadly distributed over the entire range $0<F<1$. 
The velocity bias $b_v$ is simply given by integrating 
the gaussian probability distribution $p(\delta_G| \sigma)$ over $dF$. The integrand 
is positive definite over the entire range of $F$. 
Note that the models that give the same PDF also give the same $b_v$, as shown in equation 
\ref{bFv}. We find $b_v =- 0.1 \pm 0.01$ is 
the best prediction of this model given the observed PDF. 

The bias $b_F$ has a similar behaviour for low $F$, high $\delta$ (or $\delta_G$), 
where $(\nu_2\delta +1) /( 1+\delta) \sim \nu_2$. 
However, for high $F$, corresponding to $\delta<-\nu_2^{-1}$, 
we see there is a zero crossing of $W_{b_F}$, so those regions contribute positively to bias, while the 
region $\delta>-\nu_2^{-1}$ contributes negatively to bias. 
Effects on $b_F$ are thus more complicated since the integration kernel is not negative definite everywhere: 
instead, there is a cancellation of positive and negative contributions to the integral.
Moreover, approaching $F=1$, $\delta=-1$, the contributions become large because of  $\delta+1 $ term in the denominator. 
The overall result is that $b_F$ is susceptible to the flux PDF in the voids, which is close to the continuum and so rather 
poorly determined from observations, 
suggesting the predictions for $b_F$ from flux PDF may be less reliable than predictions for $b_v$. 

It is instructive to investigate if another nonlinear transform can cure this sensitivity to void regions. 
Since we want to suppress the contribution where $F \sim 1$ the simplest example is to define a new field
$H=F-F^2/2$, whose kernel is $W_{b_H}=W_{b_F}(1-F)$ and $W_{b_{Hv}}=W_{b_{Fv}}(1-F)$. This is also shown in figure \ref{fig1}, 
and shows that it has achieved the desired effect of suppressing the contribution from the void regions. 
However, it also reduces the absolute value of the bias and velocity bias, roughly by 0.05. 
Such transformations may thus be useful if a more robust prediction of bias is needed, although they are likely 
to increase the noise. 

\subsection{Effects of scatter}

The above derivation assumes a deterministic relation between the density field and the nonlinear 
transformation. Often the relation is stochastic. 
For example, 
galaxies can be viewed as a stochastic (Poisson) sampled tracer of a nonlinear transform of the underlying density 
field and the shot noise term is added to this relation, e.g. $\delta_g=b_1\delta+b_2\delta^2+\epsilon$. Here
the scatter is additive and added after the nonlinear transformation. 
We define $\epsilon$
as a random variable uncorrelated with the density field, with $\langle \epsilon \rangle =0$ and
$\langle \epsilon^2 \rangle =\sigma_{\epsilon}^2$. 
In Lyman-$\alpha$ forest, neutral hydrogen density is a stochastic tracer of the nonlinear transformation of 
the density field, which we can model as
$\tau=\bar{\tau}(1 +\epsilon)$, with $\bar{\tau}=(1+\delta)^{\alpha}$, after which another
nonlinear transformation, $F=\exp(-\tau)$, takes place. 
We assumed a multiplicative form of scatter to avoid the unphysical situation $\tau<0$, valid as long as $\sigma_{\epsilon} \ll 1$. 
Thus depending on the model we can add the scatter either before or after the nonlinear transformation. 
If scatter is uncorrelated with the density field, $\partial \epsilon/\partial \delta_l=0$, 
then there is no contribution from the scatter to the calculation of the large scale bias above. 

However, scatter still
affects the PDF or the moments of the transformed variable 
and one must correct for this if the moments or the PDF of the transformed variable is used as a constraint. 
For example, in the case of Lyman-$\alpha$ forest we use the observed PDF to determine the bias.  
If we denote $\bar{F}=\exp(-\bar{\tau})$ as the flux in the absence of scatter and $F=\exp(-\tau)$ the observed flux with scatter, then 
since $F=\bar{F}^{1+\epsilon}$ and $dF/d\bar{F}=\bar{F}^{\epsilon}(1+\epsilon)$
we can deconvolve the observed PDF $dn/dF$ to find, 
\be
{dn \over d\bar{F}}=\int \left({dn \over dF}\right)_{F=\bar{F}(1+\epsilon)} \bar{F}^{\epsilon}\left(1+\epsilon\right) p(\epsilon | \sigma_{\epsilon})d\epsilon ,
\ee
where $p(\epsilon | \sigma_{\epsilon})=(2\pi \sigma_{\epsilon}^2)^{-1/2}\exp(-\epsilon^2/2\sigma_{\epsilon}^2)$ is the 
PDF for scatter, assumed to be gaussian. 
Simulations suggest $\sigma_{\epsilon} \sim 0.1-0.3$ at $z=3$ and 0.3-0.5 at $z=2$ \cite{1997MNRAS.292...27H}. 
The result of this analysis for $\sigma_{\epsilon}=0.4$ is shown in figure \ref{fig1}. We see that the effects on the PDF 
are modest. Once we have the deconvolved PDF we can proceed as in previous analysis. At $z=2.4$ we find the effect is to 
increase $|b_F|$ by about 5\% and $|b_v|$ by 3\%. For lower values of $\sigma_{\epsilon}$ we find even smaller effects. 

\subsection{Predictions for $b_F$ and $b_v$ from PDF of Lyman-$\alpha$ forest}

We now turn to the observational constraints on 
$b_F$, $b_v$ and $\beta$ given the observed PDF. The bias and $\beta$ are 
a function of the two parameters of 
the model, $\alpha$ and $A$. We do so by evaluating the expressions in equations \ref{bFF}-\ref{bFv}
using observed PDF in \cite{2001ApJ...549L..11M,2007MNRAS.382.1657K}. 
At $z=2.4$ we found that log-normal PDF is a reasonably good 
fit to the observed one, giving $b_{v}=-0.1$ 
Hence we expect the results to be similar to the best fit PDF in 
figure \ref{fig1}. 
This is indeed the case: we find $b_v=-0.09$ for 
\cite{2001ApJ...549L..11M} PDF and $b_v=-0.095$ for \cite{2001ApJ...549L..11M} PDF. We note that $b_v$ is independent of
the nonlinear transformation parameters. 
Bias values depend on the two parameters of the transformation. We first compare the prediction to 
the log-normal model. 
For bias we find, 
for $A=0.3$ and $\alpha=1.6$, 
 $b_F=-0.13$ and $b_F=-0.14$ 
for the two PDFs, respectively, compared to $b_F=-0.11$ for log-normal model. It is expected that 
there will be more of a difference between these and the log-normal model since in the void region 
where $F \sim 1$ the PDF is poorly measured, yet it makes a large contribution with the opposite sign. 

The amplitude $A$ is inversely proportional to UV background photoionization rate $\Gamma$,
\be
A=0.96[(1+z)/4]^{4.5}T_4^{-0.7}\Gamma_{-12}^{-1},
\label{a}
\ee
where we assumed $h=0.7$, $\Omega_m=0.27$ and $\Omega_bh^2=0.0225$ and expressed
photoionization rate $\Gamma$ in units of $10^{-12}s^{-1}$ and temperature $T$ in units of $10^4K$. 
Typical values are $\Gamma_{-12} \sim 0.5-2$ and $T_4 \sim 2$, relatively independent of redshift between 
$2<z<3$ \cite{1996ApJ...461...20H}, 
making $A$ to be increasing with redshift due to $(1+z)^{4.5}$ dependence. 
Using equation \ref{a} and the expected range of UV background amplitude 
one finds the allowed range is $0.15<A<0.6$ at $z=2.4$. 
The value that fits best the mean flux in hydrodynamic simulations is $A=0.17$ \cite{Stranex}. Hence $A=0.3$ 
is possibly too high for the observed mean flux. 
Varying the parameter $A$ while fixing $\alpha=1.6$ 
at $z=2.4$ we find $b_F\sim -0.17$ at $A=0.15$ and $b_F\sim -0.07$ at $A=0.6$ (since the differences 
between the two published PDFs is small we simply quote the average between the two). 
Varying $\alpha$ also affects the predicted vaues of bias: for $A=0.3$ we find it varies from 
$b_F\sim -0.07$ at $\alpha=1.2$ to $b_F \sim -0.18$ at $\alpha=2.0$.  

For $A=0.17$ and $\alpha=1.6$ we find $b_F\sim -0.17$ and
in combination with $b_v \sim -0.095$ this gives
$\beta \sim 0.6$. RSD $\beta$ can be as low as 0.5, since $b_F=0.2$ is at the upper end of predicted
values, and as high as 1.5, since $b_F$ can be as low as -0.06, albeit possibly at an unrealistically high $A$ or 
unrealistically low $\alpha$. 
Observed values suggest $\beta=0.8 \pm 0.2$ \cite{2011JCAP...09..001S}. 
Observations constrain best the parameter combination $b_F+b_v$  and the observed value at $z=2.4$ is $b_F+b_v=-0.30 \pm 0.01$ \cite{2011JCAP...09..001S} 
 (note that in recent literature it has become standard to divide the flux by the mean 
flux $\bar{F}$, while we use the flux itself, hence we multiply the published value by $\bar{F}$ and interpolate to 
$z=2.4$ from $z=2.25$ using the measured redshift evolution). 
At $A=0.17$, $\alpha=1.6$ our prediction is $b_F+b_v=-0.26$, slightly below the observed value. However, 
a 10\% increase of $\alpha=1.6$ or a 10\% decrease of $A=0.17$ can accommodate the observed value. 
For comparison, the linearized prediction from the term linear in $\delta$ is 
$b_F^{{\rm lin}}=A\alpha \exp{(-A)}$, which, using $\alpha=1.6$, gives -0.35 for $A=0.3$ (compared to our predicted value of -0.13) 
and -0.5 for $A=0.5$ (compared to our prediction of -0.09), while for $A=0.17$ the linearized model predicts -0.14.
We see that the linearized model grossly differs from our model for high values of $A$ where the linear approximation is inadequate, 
while for low $A$ the two are in a better agreement. 

The most robust prediction we make is for $b_v\sim -0.1$, since it does not depend on any of the transformation parameters, 
just on the PDF.
Comparing to the observations, for 
$b_F+b_v=-0.3 \pm 0.01$ and $\beta=0.8\pm 0.2$ \cite{2011JCAP...09..001S} 
we find observations suggest $b_v=-0.13\pm 0.03$ at $z=2.4$, which is within one sigma of 
our prediction. This prediction is robust in the sense that it is an analytic prediction of equation \ref{bFv} with 
no dependence on the transformation parameters, only on the flux PDF.  At this redshift the published PDFs agree with each other. 
This predicted value is lower than 
the value measured in simulations of \cite{2003ApJ...585...34M}, where $b_v \sim 0.17$ and $\beta \sim 1.6$. This discrepancy could be 
due to the simulations not matching the observed PDF, or due to additional nonlinear effects that need to be included in our model.

While the agreement between the predictions and observations is remarkably good, 
it should be pointed out that the real data contain 
absorbers with Lorentzian wings
that cannot be modeled as a simple nonlinear transformation $F=\exp(-A(1+\delta)^{\alpha})$. 
These high column density systems such as Damped Lyman $\alpha$ 
systems and Lyman limit systems increase the absorption and 
it is likely that these high density regions increase the bias (in absolute sense). Removing these regions in the data 
had almost no effect on $b_F+b_v$ \cite{2011JCAP...09..001S}, but only a small subset of these regions was identifiable in the noisy SDSS data. 
It remains an open issue how much these regions affect the bias. 
In general, only simulations can address these issues in detail. The value of our model is 
that it identifies the physical effects affecting the bias determination, while 
its quantitative predictions need to be tested in more detail against simulations and observations. 

At higher redshifts the published PDFs agree less well with each other, a consequence of the fact that the continuum 
is less well defined. For canonical value of $\Gamma_{-12}=0.5-2$ and using the two PDFs we predict $b_F \sim -0.2 \pm 0.05$ at z=3. 
The predicted values of $\beta$ are similar to $z=2.4$, around 0.6 to 0.8 in most cases although values outside this range are possible. 
There are no published values for $\beta$ from the data, but 
extrapolating the measurements of \cite{2011JCAP...09..001S} to $z=3$ one finds $b_F+b_v \sim 0.35$, compared to our 
prediction of $0.35 \pm 0.1$. 
This should be compared to the linearized prediction of $b_F \sim -0.35$ to -1. 
The conclusion from this is that the analysis gives values of bias and velocity bias in a reasonable 
agreement with published values, while the simple linearized model grossly overestimates the bias. 

The most important advantage of our model is the possibility to analytically study the influence of 
the physical parameters on the large scale bias. We find that the results are rather sensitive to small 
changes in the two model parameters. 
This is not 
surprising given the highly nonlinear nature of the transformations. It means that one can 
determine a combination of the two with the large scale bias measurements of bias. Since the two 
parameters also determine the flux PDF a combination of the two measurements provides a way to determine 
both parameters. 
In particular, assuming $A$ is determined by the mean flux then 
our analysis allows one to measure $\alpha$ from the bias, 
a complementary method to the one used in \cite{2008MNRAS.386.1131B} where the PDF was used to determine $\alpha$ 
by comparing the simulations to the observations. 
The large scale velocity bias is fully determined by the observed PDF, so this allows a robust test of 
our model predictions. Before we applying this to the data we should test our predictions against simulations, so 
a more detailed analysis of this type will be left for the future. 

\subsection{Predictions for primordial nongaussianity}

As discussed above primordial nongaussianity induces an additional bias, which can be 
scale dependent for local (equation \ref{fnl}), orthogonal or other configurations. We parametrize 
the overall effect in equation \ref{bfnl} with the bias $b_{\fnl}$. 
Equation \ref{bfnl} applied to our \lya\ model gives 
\be
b_{\fnl}=2\langle \delta {d F \over d \delta} \rangle={2(1-\alpha \beta) \over \nu_2-1}b_F.
\label{bfnllya}
\ee
We see that the prediction for $b_{\fnl}/b_F$ is determined by the values of $\alpha$ and 
$\beta$. For observed value of $\beta = 0.8$ and for $\alpha=1.6$ we find 
\be
{b_{\fnl} \over b_F} \sim -0.9.
\label{brat}
\ee
Note that in the redshift space the observed power constrains better $b_F+b_v$ \cite{2011JCAP...09..001S}, 
so a more relevant number to compare is 
\be
{b_{\fnl} \over b_F+b_v}  \sim -0.5.  
\ee

Equation \ref{brat} should be compared to the case of biased galaxies, where that ratio is given by 
${b_{\fnl} \over b}=2\delta_c(b-1)/b \sim 3.37(b-1)/b$, where $\delta_c=1.68$ (redshift 
space distortions are small for the highly biased objects).
The first thing to note is that the sign of the effect in \lya\ is negative for the chosen parameters, while the 
effect is positive for biased tracers with $b>1$: primordial nongaussianity with positive 
$\fnl$ supresses power in \lya. 
We also see 
that the relative effect of primordial nongaussianity in \lya\ is significantly reduced 
relative to the highly biased tracers, since 0.5 or 0.9 is much less than 3.37. 
This prediction in fact sensitively depends on the parameters $\alpha$ and $\beta$, as seen in equation \ref{bfnllya}.
The primordial nongaussianity effect vanishes if  $\beta = \alpha^{-1}$, since this is the same as the linearized limit 
$F^{{\rm lin}}=A\alpha \exp{(-A)}\delta$,
where the primordial nongaussianity signatures do not appear in 2-point correlations. 
The effect changes sign for $\beta < \alpha^{-1}$ relative to the sign above.
On the other hand, it should be pointed out that the overall effect scales by the inverse of the growth 
rate $D(z)$ (equation \ref{eq:alphaparam}), which for \lya\ redshifts can be a factor of 2-3 larger relative to $z=0$. 

Since the predictions for \lya\ are sensitive to the parameters and very close to zero it is worth 
exploring if a further nonlinear transformation of \lya\ can change the nongaussian bias relative to gaussian bias. For example, 
for $H=F-F^2/2$ one finds $b_H=-0.07$, $b_v=-0.04$ and $b_{\fnl}=-0.04$, so that $b_{\fnl H}/b_H=0.6$, 
reversing the sign of the effect. Similarly, defining a new field as $G=(F-\bar{F})^2$ gives $b_G=2[(1-\bar{F})b_F-b_H]$,
so that for $\bar{F}=0.8$ one has $b_{\fnl G}/b_G=1.2$, which is also positive in sign and a factor of 2 larger.
If we want a stronger contrast we can arrange the bias to vanish. For example, 
for $K=0.6F-H$ we find $b_{\fnl K}/b_K=\infty$, because the large scale bias $b_K$ vanishes for 
this combination. We can also choose a different combination such that $\beta$ or $b(1+\beta)$ vanishes 
(but we cannot make all the large scale power in redshift space vanish since we cannot simultaneously vanish 
density and velocity bias). Still, such transformations may boost the primordial nongaussianity signal relative to the 
gaussian signal on large scales. This is related to the multi-tracer method of canceling sampling variance \cite{2009PhRvL.102b1302S}, where two tracers 
with different bias values are combined in a way that the large scale fluctuations in the density field cancel: this in fact happens 
if one considers the linear combinations of the two tracers where the large scale bias vanishes, in which case one 
is left with the large scale primordial nongaussianity bias, which does not vanish. 
Such nonlinear transforms do not necessarily enhance the signal itself and 
typically increase the noise properties of the new field, so only in the sampling variance limit one gains using the 
sample variance canceling techniques. We do not expect to be in such a limit for \lya\ with current data sets \cite{2007PhRvD..76f3009M,2011MNRAS.415.2257M}. 

Yet another potential advantage of these nonlinear transformations is to test against systematics. On large scales there are 
numerous sources of additional power in \lya\ such as UV background fluctuations or He reionization signatures 
\cite{2005MNRAS.360.1471M,2011MNRAS.415..977M}, which could mimic the scale dependent bias effects of primordial nongaussianity. However, primordial nongaussianity changes in a predictable 
way under the nonlinear transformation. For example, if one observes an unexplained large scale power in \lya\ compatible 
with primordial nongaussianity one can devise a nonlinear transformation where primordial nongaussianity 
vanishes, while one would not expect the rest of the effects to vanish.  
A more detailed analysis to see how much can be gained by such methods is beyond the scope of this paper. 

\section{Conclusions}

In this paper we discuss the large scale clustering properties of 
nonlinear transformations of density field. 
We show that on large scales these can be viewed as biased 
versions of the density field itself. 
We present an analytic derivation of bias, velocity bias 
and 
primordial nongaussianity bias.
The interesting aspect of our calculation is that the large scale bias can be expressed entirely in terms of the 
final PDF and the physical parameters of the nonlinear transformation, allowing one to determine it from the 
observed PDF without the need for simulations. 
The resulting bias depends on the nature of the transformation and 
with a suitable nonlinear transformation 
one can design a field with very different large scale bias 
than the original field, including bias of zero.

The primary application in this paper is to the ionizing equilibrium model of Lyman-$\alpha$ forest: 
we derive the bias, velocity bias and primordial nongaussianity bias of \lya\
flux as a function of redshift from the observed PDF. 
Velocity bias has a very simple expression (equation \ref{bFv}) that 
depends only on the observed PDF and its predicted value is in a good agreement with observations of \cite{2011JCAP...09..001S}. 
We derive the bias as a function of parameters of transformation, which are dominated by the UV background amplitude and the
slope of the temperature-density relation. 
Assuming their fiducial values the values of bias and $\beta$ are in a reasonable agreement with observations of \cite{2011JCAP...09..001S}. 
We find that the bias has a significant contribution from the void regions, where lognormal PDF
differs from the measured one, as seen in figure \ref{fig1} . Another nonlinear transformation, e.g. to 
$H=F-F^2/2$, can reduce the sensitivity to voids. We find that the primordial nongaussianity bias for the observed $\beta \sim 0.8$ 
has the opposite sign than highly biased galaxy tracers, but is relatively small. A small change in the 
parameters can change its sign or make it zero. 

There are several generalizations and 
applications of the model that can be pursued. First, we used the linear growth evolution to derive the coupling between 
long and short wavelength modes. This may break down when $\sigma_{\delta}^2=\langle \delta^2 \rangle \gg 1$. 
It would be useful to extract from simulations the general value of coefficients $d \delta^n/d\delta_l$
as a function of scale, and use these in the calculations above. 
With this one should be able to fully compute the bias for any nonlinear transformation and connect 
it to its PDF. 
One can also extend the calculation to the next order and investigate second order bias, 
$b_2=(\partial^2 \tau / \partial \delta_l^2)/2$. For example, one could design a transformation where 
second order bias vanishes, therefore suppressing the scale dependence of the bias. 
Similarly, one can also design a transformation where velocity bias vanishes, making the redshift space 
correlation function isotropic on large scales. 
Or, one could design a transformation where the bias vanishes, in which case all the large scale correlations 
would come from velocities. 

Having established that a nonlinear transformation traces the large scale modes, but with a different bias, 
another interesting question worth further investigation
is whether such transformations can increase the signal to noise (S/N) of the power spectrum measurement.  
On large scales this is proportional to $b^2/\sigma^2$, where $\sigma^2$ is the variance of the field, including 
observational noise. 
While an increase in bias 
increases the signal, the accompanying nonlinear transformation may also increase the noise, so it is not clear if
$S/N$ can increase. For example, taking a log of flux $F$ gives optical depth $\tau=-\ln F$, which for $\alpha=2$ has 
$b_{\delta \tau} \sim 2-3$ (equation \ref{btau}), compared to $b_{F} \sim -0.15$, a huge increase in bias. 
However, the corresponding nonlinear transformation also greatly enhances the noise, both detector and photon noise, 
specially for $F \sim 0$, as well as Poisson like noise caused by high neutral hydrogen column density sources in the 
optical depth, which makes this example likely to be impractical. 
There may however be other nonlinear transformations that perform better. 
A detailed analysis is beyond the scope of this paper. 

Another application of the method is to multi-tracer methods: 
in \cite{2009JCAP...10..007M,2009PhRvL.102b1302S} it was argued that combining two tracers with different bias parameters can reduce the
sampling variance on certain quantities, such as the RSD $\beta$ or nongaussianity parameter $f_{nl}$. 
This is because if the two tracers both trace the underlying long wavelength $\delta_l$, taking the 
ratio of the two eliminates $\delta_l$ itself, and so eliminates the dominant source of error on large 
scales, which is the stochastic nature of $\delta_l$ (i.e., sampling variance). At the same 
time this ratio contains some useful cosmological information, such as sensitivity to primordial nongaussianity \cite{2009PhRvL.102b1302S}. 
Here we have shown that the second tracer with a different bias can be simply 
obtained by a nonlinear transformation of the field itself and that takig a linear combination of the two can result in 
vanishing of the large scale bias, effectively achieving the sampling variance cancellation. 
These methods are of course simply reduced versions of combining N-point statistics, but are particularly simple 
to understand and analyze. 
In this view the nonlinear transformation converts the higher order correlations 
of the density field into the 2-point correlations of the nonlinear transform.
For example, if we square the density field and correlate with the density field itself we 
obtain a reduced version of a 
3-point function, which we can then compare to the 2-point function of the density field. Both will trace the 
long wavelength modes on large 
scales, but the first one will be biased relative to the density field itself. 
Comparing the two thus eliminates the sampling variance, while preserving some cosmological information in the 
bias itself: for example, its sensitivity to primordial local 
nongaussianity has been shown in e.g. \cite{2011PhRvD..84h3509H}. 
While this example involves comparing 2 and 3 point functions, other combinations may be more effective in 
extracting the information optimally.  

Testing the results against systematics may be another application of such nonlinear transforms. This is specially 
true for primordial nongaussianity, which enhances the power on very large scales. Other effects may achieve the 
same effect: for \lya\ the additional enhancement of power could come from UV background fluctuations or He reionization, 
for galaxies the large scale power could be enhanced due to the effects from our own galaxy, such as star-galaxy separation or extinction. 
In most cases it may be difficult to separate these effects from the primordial nongaussianity. 
However, one can use a nonlinear tranform to change the primordial nongaussianity bias and using the expressions derived here 
we can predict what their large scale bias is. For example, as discussed in this paper
one can devise a noninear transform where primordial nongaussianity bias vanishes, while one would not expect the rest of the 
effects to vanish, so one can separate the two effects. 

In summary, the results of this paper can be used to
develop an analytic understanding of the large scale bias of nonlinear transformations 
and its sensitivity to parameters of the transformation. 
This is required if we want to achieve a better theoretical understanding of 
bias of \lya\ and other nonlinear tracers of matter density. 
Our results allow one to design nonlinear transformations with a nearly arbitrary large scale bias, 
which can be used to optimize the extraction of cosmological information 
from the cosmological observations. 

This work is supported by the DOE, the Swiss National Foundation under contract 200021-116696/1 and WCU grant R32-10130.
I thank Pat McDonald and Matteo Viel for providing Lyman-$\alpha$ forest PDFs in data form. I also thank 
Tobias Baldauf, Vincent Desjacques, Andreu Font, Nico Hamaus, Shirley Ho, Pat McDonald and An\v ze Slosar for useful discussions. 

\bibliography{cosmo,cosmo_preprints}
\bibliographystyle{revtex}

\end{document}